\documentclass[11pt, a4paper]{article}

\usepackage[utf8]{inputenc}
\usepackage[T1]{fontenc}
\usepackage{amsmath, amssymb, amsthm}
\usepackage{graphicx}
\usepackage[round]{natbib} 
\usepackage{booktabs}
\usepackage{hyperref}
\usepackage{geometry}
\title{Exact Comparison of Explanatory Strength of Two Dependent Predictors}
\author{
    Tomáš Mrkvička \\
    \small University of South Bohemia, Faculty of Agriculture and Technology \and 
    Jan Radimský \\
    \small University of South Bohemia, Faculty of Arts
}
\date{\today}

\begin{document}

\maketitle

\begin{abstract}
Comparing the relative explanatory power of two dependent predictors regarding a common target variable is a fundamental challenge across scientific disciplines. Classical asymptotic procedures, such as Vuong's closeness test or the Hotelling-Williams test, frequently collapse under pathological data conditions, including heavy-tailed distributions and extreme categorical sparsity. To bypass these limitations, practitioners often turn to non-parametric resampling. However, naive permutation tests destroy the natural covariance structure of dependent predictors, while the paired bootstrap—evaluating variance around the alternative hypothesis and introducing artificial ties—suffers from metric space compression and categorical omission, rendering it highly unreliable in finite samples. 

In this paper, we introduce the \textit{Paired Swap Permutation Test}, a novel and exact non-parametric methodology. Grounded in the principle of functional exchangeability under the null hypothesis, our algorithm utilizes a symmetric within-subject swapping mechanism for categorical data, and introduces an Empirical Cumulative Distribution Function (ECDF) mapping step for continuous domains. This copula-based transposition perfectly preserves marginal densities and the empirical support without introducing resampling ties. 

Through extensive Monte Carlo simulations, we demonstrate that the proposed test strictly maintains the nominal significance level and maximizes statistical power under conditions where standard methods become catastrophically liberal or pathologically conservative. Finally, we apply the framework to a high-dimensional linguistic dataset of Italian noun-noun compounds, proving its capacity to deliver robust, exact inference in environments where conventional analytical methods inherently fail.
\end{abstract}

\section{Introduction}
\label{sec:introduction}

In many scientific disciplines, researchers are tasked with determining which of two available predictors better explains the variance or distribution of a target variable. Formally, given a response variable $Y$ and two dependent predictors $X_1$ and $X_2$ measured on the same set of observations, the objective is to test the null hypothesis that $X_1$ and $X_2$ possess equal explanatory strength regarding $Y$. This problem naturally arises in diverse applied fields. 

In linguistics, for instance, a recurrent question is whether the first or the second constituent lemma of a compound word serves as a stronger predictor of the semantic or grammatical properties of the resulting compound. Related questions have been investigated in research on compound headedness, conceptual combination, constituent family size, and compound processing, where the relative contribution of individual constituents to compound interpretation has been repeatedly examined. In particular, it has been argued that the semantic relationship between compound constituents may be associated more strongly with one constituent than with the other \citep{gagne_influence_1997,gagne_constituent_2009,gunther_caoss_2021}. Furthermore, the family size of constituents (i.e., the number of compounds in which a particular constituent appears) has been shown to affect compound interpretation and processing \citep{de_jong_morphological_2000}. However, these studies typically focus on semantic transparency, interpretation, productivity, or processing effects rather than on the formal statistical comparison of the explanatory strength of two dependent predictors. This is partly because they primarily investigate Noun+Noun compounds in languages such as English or German, where the Noun+Noun compounding pattern is highly productive and allows the relatively unrestricted formation of new compounds. In Romance languages, by contrast, the Noun+Noun compounding pattern represents a more recent innovation whose spread seems to be driven by recurring lexically specified constituents \citep{radimsky_paradigmatic_2020}. Consequently, the relationship between compound constituents and compound classification poses distinct theoretical and methodological challenges and provides an opportunity for methodological innovation. More generally, it exemplifies a broader statistical problem: comparing the explanatory strength of two dependent predictors measured on the same observations.

Traditionally, this problem is addressed using analytical and asymptotic models. For continuous domains, dependent correlation tests such as the Hotelling-Williams \cite{hotelling1940selection, williams1959comparison} or Steiger's $Z$-test \cite{steiger1980tests} are widely employed. However, these methods are strictly limited to linear dependencies (Pearson's correlation) and assume multivariate normality, causing them to fail catastrophically when applied to heavy-tailed distributions or rank-based/non-linear dependence measures (e.g., Kendall's $\tau$, distance covariance). In categorical domains, practitioners frequently rely on Vuong's closeness test for non-nested models \cite{vuong1989likelihood}. Yet, Vuong's test relies on asymptotic likelihood ratios that systematically break down in the presence of data sparsity—a common linguistic pathology where perfect separation or zero-frequency cells cause the test to collapse due to overfitting.

The limitations of asymptotic approximations inherently lead researchers to non-parametric resampling techniques. While seemingly straightforward, resampling with dependent predictors is frequently mishandled in applied research. Standard permutation tests require the assumption of exchangeability. However, practitioners often resort to "naive" permutation tests that independently shuffle $X_1$ and $X_2$. As previously established in literature regarding random forests \cite{strobl2008conditional} and genomics \cite{abney2015permutation}, independent shuffling of dependent variables destroys their natural covariance structure. This artificially deflates the variance of the null distribution, leading to invalid $p$-values and a severe loss of statistical power. We intentionally include naive permutations in our simulation studies as an analytical baseline to demonstrate the massive scale of Type I errors generated when shared information is ignored.

To avoid destroying the covariance structure, a common alternative is the paired bootstrap, which resamples observation rows with replacement \cite{efron1993introduction}. Unfortunately, the paired bootstrap evaluates the variance around the local observed alternative hypothesis ($H_1$) rather than the strict null ($H_0$). Furthermore, sampling with replacement inevitably introduces artificial ties (duplicate observations). We will demonstrate that, depending on the chosen test statistic and the underlying data pathologies, the paired bootstrap can become either pathologically conservative or artificially liberal.  Ultimately, these findings underscore that for dependent predictors in finite samples, the paired bootstrap remains a purely approximate method highly vulnerable to inferential instability.

The purpose of this paper is to bridge the gap between classical asymptotic theory and exact non-parametric inference. We introduce a novel, robust permutation methodology - the \textit{Paired Swap Permutation Test}. Instead of independent shuffling or resampling with replacement, our algorithm constructs the exact null distribution directly under $H_0$ by iteratively swapping the functional roles of $X_1$ and $X_2$ within the observation vector. Because the test statistics are functionally exchangeable under the null hypothesis, this procedure provides an exact test \citep{pesarin2010permutation,ernst2004permutation}. While paired permutation tests are well-established in the literature for comparing the location parameters of two variables, our approach re-purposes the swapping mechanism to solve a fundamentally different problem: evaluating the relative predictive strength of two dependent covariates regarding a third target variable. The core novelty of the \textit{Paired Swap Permutation Test} lies in its application to asymmetric continuous domains; by introducing an Empirical Cumulative Distribution Function (ECDF) mapping step, our algorithm enables the exchange of relative ranks between predictors with differing marginal distributions, thereby ensuring exact exchangeability while perfectly preserving the original marginal densities.

Although standard computational environments and R packages (such as \texttt{coin}) provide robust frameworks for conditional inference procedures, they currently lack support for comparing non-nested dependent predictors via non-linear estimators, particularly when their marginal distributions differ. To bridge this methodological void, we introduced the \textit{Paired Swap Permutation Test}. To ensure immediate accessibility, transparency, and computational reproducibility, the complete \texttt{R} implementation of our proposed algorithm—including the ECDF-mapping procedure and the scripts used for all simulation studies is provided as open-source supplementary material hosted on GitHub \url{https://github.com/mrkvickatoma-cloud/Swap_Permutation}.

The remainder of this article is organized as follows. Section \ref{sec:theoretical_framework} formalizes the statistical problem and reviews established test statistics and asymptotic models for both continuous and discrete domains. Section \ref{sec:resampling} outlines existing non-parametric resampling strategies, details their theoretical shortcomings in the context of dependent predictors, and formally introduces our proposed \textit{Paired Swap Permutation Test}. Section \ref{sec:simulation} presents an extensive Monte Carlo simulation study that benchmarks our exact method against classical and approximate techniques across various data pathologies. Section \ref{sec:case_study} demonstrates the practical necessity and robust performance of the proposed framework through an empirical application to Italian noun-noun compounds. Finally, Section \ref{sec:conclusion} provides a concluding discussion of our findings and their broader implications for applied research.

\section{Problem Formulation and Theoretical Framework}
\label{sec:theoretical_framework}

Let $Y$ be a target random variable and let $X_1$ and $X_2$ be two dependent predictors measured on the same set of $N$ observations. The core objective is to determine which predictor contains more information about $Y$. Formally, let $S(\cdot, \cdot)$ denote a generic measure of predictive strength or statistical association. We frame our objective as testing the null hypothesis of equal explanatory power:
\begin{equation}
    H_0: S(X_1, Y) = S(X_2, Y)
\end{equation}
against the alternative hypothesis $H_1: S(X_1, Y) \neq S(X_2, Y)$. The choice of the function $S$ depends fundamentally on the distributional properties of the data domain. In this section, we review the established test statistics and their asymptotic counterparts for both continuous and discrete scenarios.

\subsection{Continuous Domains}

When evaluating relationships in continuous metric spaces, the explanatory strength is conventionally measured via correlation coefficients. 

To test $H_0$ asymptotically under linear assumptions, researchers typically rely on Hotelling's adaptation of the $t$-test, later refined by \citet{williams1959comparison}. Let $r_{y1}$, $r_{y2}$, and $r_{12}$ denote the sample Pearson correlations between the respective variables. The Hotelling-Williams test statistic evaluates the difference $(r_{y1} - r_{y2})$ while explicitly penalizing for the cross-correlation $r_{12}$ between the predictors:
\begin{equation}
    t = (r_{y1} - r_{y2}) \sqrt{\frac{(N - 3)(1 + r_{12})}{2 (1 - r_{y1}^2 - r_{y2}^2 - r_{12}^2 + 2r_{y1}r_{y2}r_{12})}}
\end{equation}
Under the assumption that $(Y, X_1, X_2)$ follows a multivariate normal distribution, this statistic follows a Student's $t$-distribution with $N - 3$ degrees of freedom. However, this classical approach is strictly confined to linear relationships and is highly vulnerable to non-normality. To robustly capture diverse types of dependencies, our study examines three distinct test statistics $S$:

\begin{enumerate}
    \item \textbf{Pearson's correlation ($r$):} Used when the relationship is strictly linear. While computationally efficient, it relies heavily on finite variance assumptions and is notoriously susceptible to variance explosion in the presence of outliers or heavy-tailed distributions.
    \item \textbf{Kendall's rank correlation ($\tau$):} A non-parametric, rank-based alternative used for capturing monotonic relationships. Because it operates on the topological ranks rather than metric distances, it exhibits strong robustness against heavy tails. \item \textbf{Distance Covariance (dcov):} Introduced by \citet{szekely2007measuring}, distance covariance is an energy-based statistic capable of detecting arbitrary (both linear and non-linear) dependencies. It equals zero if and only if the variables are strictly independent. 
\end{enumerate}

Because the Hotelling-Williams asymptotic framework cannot be applied to Kendall's $\tau$ or distance covariance, comparing these advanced statistics relies inherently on resampling methods.

\subsection{Discrete Domains}

In categorical data analysis, such as the linguistic evaluation of compound constituents, variables occupy discrete state spaces. For our exchangeability framework to be theoretically sound, we restrict our focus to scenarios where $X_1$ and $X_2$ share the exact same categorical state space (i.e., both predictors are drawn from the same predefined vocabulary or set of classes).

For discrete distributions, the classical asymptotic approach for model selection is Vuong's closeness test \citep{vuong1989likelihood}. From an information-theoretic perspective, Vuong's test evaluates the expected Kullback-Leibler (KL) divergence between two candidate models and the true data-generating process. Let $\hat{p}(y_i|x_{1i})$ and $\hat{p}(y_i|x_{2i})$ represent the estimated conditional probabilities of observing the target outcome $y_i$ given the predictors $x_{1i}$ and $x_{2i}$, respectively. Vuong's test is based on the pointwise log-likelihood ratios:
\begin{equation}
    m_i = \log \left( \frac{\hat{p}(y_i | x_{1i})}{\hat{p}(y_i | x_{2i})} \right)
\end{equation}
The standardized test statistic is then constructed as:
\begin{equation}
    V = \frac{\sqrt{N} \bar{m}}{s_m}
\end{equation}
where $\bar{m}$ is the sample mean and $s_m$ is the sample standard deviation of the pointwise ratios $m_i$. Under the null hypothesis that both predictors provide equally accurate probabilistic models for $Y$, the statistic $V$ converges in distribution to a standard normal $\mathcal{N}(0,1)$.

While mathematically rigorous under ideal conditions, Vuong's test relies entirely on valid likelihood estimates and asymptotic normality. In applied categorical scenarios—especially in natural language processing—data is frequently characterized by extreme sparsity. In finite sparse samples, empty cells and perfect separation cause estimated probabilities to approach zero. This causes the log-likelihood ratios $m_i$ to diverge to infinity, leading to a catastrophic breakdown of the variance estimate $s_m$ and the entire asymptotic approximation.

To overcome the limitations of likelihood-based asymptotics, we utilize \textbf{Mutual Information} (also known as Information Gain) as our non-parametric test statistic $S$. Mutual information $I(X; Y)$ quantifies the reduction in uncertainty about the target $Y$ obtained by observing the predictor $X$:
\begin{equation}
    I(X; Y) = \sum_{x \in \mathcal{X}, y \in \mathcal{Y}} p(x, y) \log \left( \frac{p(x, y)}{p(x)p(y)} \right)
\end{equation}
The null hypothesis is thus reformulated as $H_0: I(X_1; Y) = I(X_2; Y)$. Because the marginal entropy of the target $H(Y)$ is constant across both comparisons, maximizing mutual information is mathematically equivalent to minimizing conditional entropy. However, framing the statistic as "information explained" provides the exact discrete analogue to the squared correlation coefficient in continuous domains. Mutual information is exceptionally advantageous for discrete comparisons: it makes no assumptions about ordinality, natively handles arbitrary non-linear categorical mappings, and provides a direct, interpretable measure of predictive strength without relying on unstable likelihood ratios in sparse contingency tables.

\section{Resampling Strategies}
\label{sec:resampling}

When asymptotic approximations fail due to finite sample sizes, heavy-tailed distributions, or data sparsity, non-parametric resampling provides a necessary alternative. In this section, we first define the general Monte Carlo testing framework used to evaluate differences in explanatory power. We then review standard resampling practices—namely naive permutations and the paired bootstrap—and highlight their theoretical shortcomings in the context of dependent predictors. Finally, we formally introduce the \textit{Paired Swap Permutation Test}, an exact methodology designed to overcome these limitations.

\subsection{The General Monte Carlo Testing Framework}
Regardless of the specific resampling algorithm employed, all non-parametric methods in this study share a common Monte Carlo evaluation structure. 

Let $S(\cdot, \cdot)$ be the chosen measure of explanatory strength (e.g., Pearson's $r$, Kendall's $\tau$, distance covariance, or Mutual Information). The observed test statistic is defined as the empirical difference in explanatory power between the two predictors:
\begin{equation}
    T_{obs} = S(X_1, Y) - S(X_2, Y)
\end{equation}

The testing procedure consists of generating $B$ resampled or permuted datasets. In each iteration $b \in \{1, \dots, B\}$, a new pair of predictor vectors $(X_1^{(b)}, X_2^{(b)})$ is generated according to a specific resampling strategy, while the target vector $Y$ typically remains fixed. The test statistic is then re-evaluated on the permuted data:
\begin{equation}
    T^{(b)} = S(X_1^{(b)}, Y) - S(X_2^{(b)}, Y)
\end{equation}

Finally, the two-sided empirical $p$-value is calculated by comparing the observed statistic against the constructed null distribution. To ensure a valid test that does not produce a $p$-value of exactly zero, the observed sample is included as one of the permutations:
\begin{equation}
    p = \frac{1 + \sum_{b=1}^{B} I\left(|T^{(b)}| \ge |T_{obs}|\right)}{B + 1}
\end{equation}
where $I(\cdot)$ is the indicator function. The fundamental difference between competing non-parametric methods lies entirely in the algorithm used to generate the surrogate variables $X_1^{(b)}$ and $X_2^{(b)}$.

\subsection{The Naive Permutation Approach}
A common practice in many applied fields is the use of "naive" permutation tests. Under this approach, the surrogate variables are generated by randomly and independently shuffling the original predictor vectors: $X_1^{(b)} = \pi_1(X_1)$ and $X_2^{(b)} = \pi_2(X_2)$, where $\pi_1$ and $\pi_2$ are independent random permutations.

As briefly discussed in Section \ref{sec:introduction}, this approach fundamentally violates the assumption of exchangeability when predictors are dependent. Because $Y$, $\pi_1(X_1)$ and $\pi_2(X_2)$ are unlinked, the independent shuffling effectively destroys the natural covariance structure (the shared information) between the predictors. Consequently, the empirical variance of $T^{(b)}$ is artificially distorted. As rigorously demonstrated in literature regarding random forests \citep{strobl2008conditional} and genomic associations \citep{abney2015permutation}, ignoring this dependence leads to invalid $p$-values.

\subsection{The Paired Bootstrap Approach}
To preserve the covariance structure, the standard alternative is the paired bootstrap \citep{efron1993introduction}. Instead of permuting vectors independently, the bootstrap generates the $b$-th dataset by sampling $N$ row indices $J^{(b)} = \{j_1, \dots, j_N\}$ from the set $\{1, \dots, N\}$ with replacement. The surrogate variables are thus $(Y^{(b)}, X_1^{(b)}, X_2^{(b)}) = (Y_J, X_{1J}, X_{2J})$.

While this accurately maintains the joint dependency structure, it suffers from severe limitations in finite or sparse samples. The fundamental bias of the bootstrap lies in its evaluation of variance. Because it resamples data from the empirical distribution (which reflects the alternative hypothesis $H_1$), the values of $T^{(b)}$ estimate the variability at a point in the information space that may radically differ from the true point of equivalence ($H_0$). Given that information and correlation metrics are strictly bounded and heteroscedastic, this $H_1$-based variance systematically distorts the null distribution.

Furthermore, sampling with replacement inherently produces datasets containing duplicate observations while omitting others. In categorical domains under sparsity, rare categories are systematically dropped, artificially compressing the entropy of $X^{(b)}$. 

\subsection{The Paired Swap Permutation Test}
To conduct exact non-parametric inference without violating the dependence structure or compressing the data space, we propose the \textit{Paired Swap Permutation Test}. Our framework is grounded in the principle that under the null hypothesis ($H_0$), $X_1$ and $X_2$ provide equivalent explanatory power. Because they are functionally symmetric in their predictive capacity, their roles within the observation vector are interchangeable. By generating $X_1^{(b)}$ and $X_2^{(b)}$ strictly under $H_0$, the statistics $T^{(b)}$ remain functionally exchangeable, mathematically guaranteeing an exact $p$-value \citep{pesarin2010permutation}.

\subsubsection{Categorical Domains: The Symmetric Swap}
Let $\mathcal{D} = \{(y_i, x_{1i}, x_{2i})\}_{i=1}^N$ be a dataset of $N$ paired observations. In categorical domains (e.g., lexical constituents in linguistics), $X_1$ and $X_2$ frequently share the exact same state space. Under $H_0$, their conditional probability distributions are strictly equivalent: $P(Y | X_1) = P(Y | X_2)$.

Because the predictors are equivalent in their explanatory power, rather than independently shuffling the columns or resampling with replacement, our procedure constructs the surrogate variables $X_1^{(b)}$ and $X_2^{(b)}$ by iteratively swapping the predictor values \textit{within} the same observation.
For each permutation $b \in \{1, \dots, B\}$, we generate a random binary mask vector $\mathbf{m}^{(b)} \in \{0, 1\}^N$, where each element is drawn from a Bernoulli distribution with $p = 0.5$.

The elements of the permuted predictor vectors for iteration $b$ are defined as:
\begin{equation}
    x_{1i}^{(b)} = 
    \begin{cases} 
      x_{1i} & \text{if } m_i^{(b)} = 0 \\
      x_{2i} & \text{if } m_i^{(b)} = 1 
   \end{cases}
   \quad \text{and} \quad
    x_{2i}^{(b)} = 
    \begin{cases} 
      x_{2i} & \text{if } m_i^{(b)} = 0 \\
      x_{1i} & \text{if } m_i^{(b)} = 1 
   \end{cases}
\end{equation}

This within-subject swapping mechanism maintains the exact marginal distributions of the combined predictor space while perfectly preserving the correlation structure. The statistic $T^{(b)}$ in this case, the difference in Mutual Information $\Delta I^{(b)}$ is evaluated safely without the risk of sparsity-induced entropy compression.

\subsubsection{Continuous Domains: The ECDF-Mapped Swap}
Extending the swap principle to continuous domains introduces a critical challenge: $X_1$ and $X_2$ may possess entirely different marginal distributions, e.g., a normal versus a log-normal distribution. A direct raw-value swap would violate the marginal densities, effectively testing for differences in distributional shape rather than isolating predictive strength.

To resolve this, we introduce an Empirical Cumulative Distribution Function (ECDF) mapping step, acting as a discrete Probability Integral Transform (PIT). Instead of exchanging raw values, we exchange the \textit{relative distributional ranks}.

Let $R(x_{1i})$ and $R(x_{2i})$ denote the ranks of the $i$-th observation within their respective sorted ascending vectors $\mathbf{X}_1^{\uparrow}$ and $\mathbf{X}_2^{\uparrow}$. If the random mask dictates a swap ($m_i^{(b)} = 1$), the values are substituted across domains by matching their ECDF quantiles:
\begin{equation}
\begin{aligned}
    x_{1i}^{(b)} &= \mathbf{X}_1^{\uparrow} \left[ R(x_{2i}) \right] \\
    x_{2i}^{(b)} &= \mathbf{X}_2^{\uparrow} \left[ R(x_{1i}) \right]
\end{aligned}
\end{equation}

This ECDF-mapped swap ensures that any extreme value (outlier) in $X_1$ retains its relative extremity when hypothetically transposed into the domain of $X_2$. By exchanging relative ranks rather than raw values, the algorithm strictly prevents the injection of out-of-domain values that would severely corrupt the marginal density profiles, e.g., inadvertently mixing normally and exponentially distributed predictors.

\section{Simulation Study}
\label{sec:simulation}

To empirically validate the theoretical properties of the \textit{Paired Swap Permutation Test}, we conducted an extensive simulation study. The objective is to benchmark our proposed method against asymptotic standards, the naive permutation test, and the paired bootstrap. We evaluate the empirical Type I error rate (under $H_0$) and statistical power (under $H_1$) across two principal axes: the degree of shared information (cross-correlation $\rho$) between the predictors, and the available sample size ($N$). 

\subsection{Simulation Design}

The simulations were executed in two distinct domains to reflect the diverse applications of the proposed methodology. All empirical rejection rates were computed at a nominal significance level of $\alpha = 0.05$ over $B = 500$ resampling iterations and $2000$ Monte Carlo replications per condition.

\subsubsection{Continuous Domain}
To rigorously stress-test the continuous estimators, we constructed a data-generating process heavily reliant on non-normality and asymmetric marginal densities. The underlying dependence structure was generated via a Gaussian copula. We simulated a latent multivariate normal vector $(Z_Y, Z_1, Z_2) \sim \mathcal{N}(0, \Sigma)$, where the covariance matrix $\Sigma$ controlled the true explanatory strength and the shared information $\rho$.

The latent vectors were transformed into the uniform domain via the standard normal CDF and subsequently mapped to the following highly asymmetric empirical distributions:
\begin{itemize}
    \item \textbf{Target $Y$:} Log-normal distribution $LN(0, 1.5)$, introducing a severe heavy right tail to intentionally violate linear asymptotic assumptions.
    \item \textbf{Predictor $X_1$:} Normal distribution $\mathcal{N}(10, 2)$.
    \item \textbf{Predictor $X_2$:} Log-normal distribution $LN(2, 0.5)$.
\end{itemize}
Under the null hypothesis ($H_0$), the latent correlations with the target were symmetric ($r_{y1} = r_{y2} = 0.5$). For the power analysis ($H_1$), an asymmetric predictive advantage was introduced ($r_{y1} = 0.6$, $r_{y2} = 0.3$). We evaluated performance across a sequence of shared correlations $\rho \in [0, 0.9]$ at a fixed sample size of $N=30$, and across varying sample sizes $N \in [10, 100]$ at a fixed $\rho=0.5$.

\subsubsection{Discrete Domain}
For categorical data, we simulated a linguistic-style compound structure where\\ $Y \in \{\text{Type A, Type B, Type C}\}$ and both predictors share the identical vocabulary $X_1, X_2 \in \{x_1, \dots, x_{10}\}$. The dependence was governed by a latent categorical root variable $Z$.

Under $H_0$, with probability $\rho$, both predictors inherited the exact same latent root $Z$ (shared information), and with probability $1-\rho$, they were drawn from independent latent roots. Localized noise was then injected by shifting the predictor value by $\{-1, 0, +1\}$ within the categorical space. Because the noise magnitude was identical, $X_1$ and $X_2$ possessed equivalent Mutual Information regarding $Y$. 
Under $H_1$, an asymmetric noise profile was applied: $X_1$ strictly retained the true latent root without noise (strong predictor), whereas $X_2$ was subjected to severe noise spanning $\{-2, \dots, +2\}$ (weak predictor). Similar to the continuous case, simulations were mapped across $\rho \in [0, 0.9]$ ($N=30$) and $N \in [20, 100]$ ($\rho=0.5$).

\subsection{Results}

\subsubsection{Continuous Evaluation: Pearson and Distance Covariance}
\begin{figure}[htpb]
    \centering
    \includegraphics[width=7.5cm]{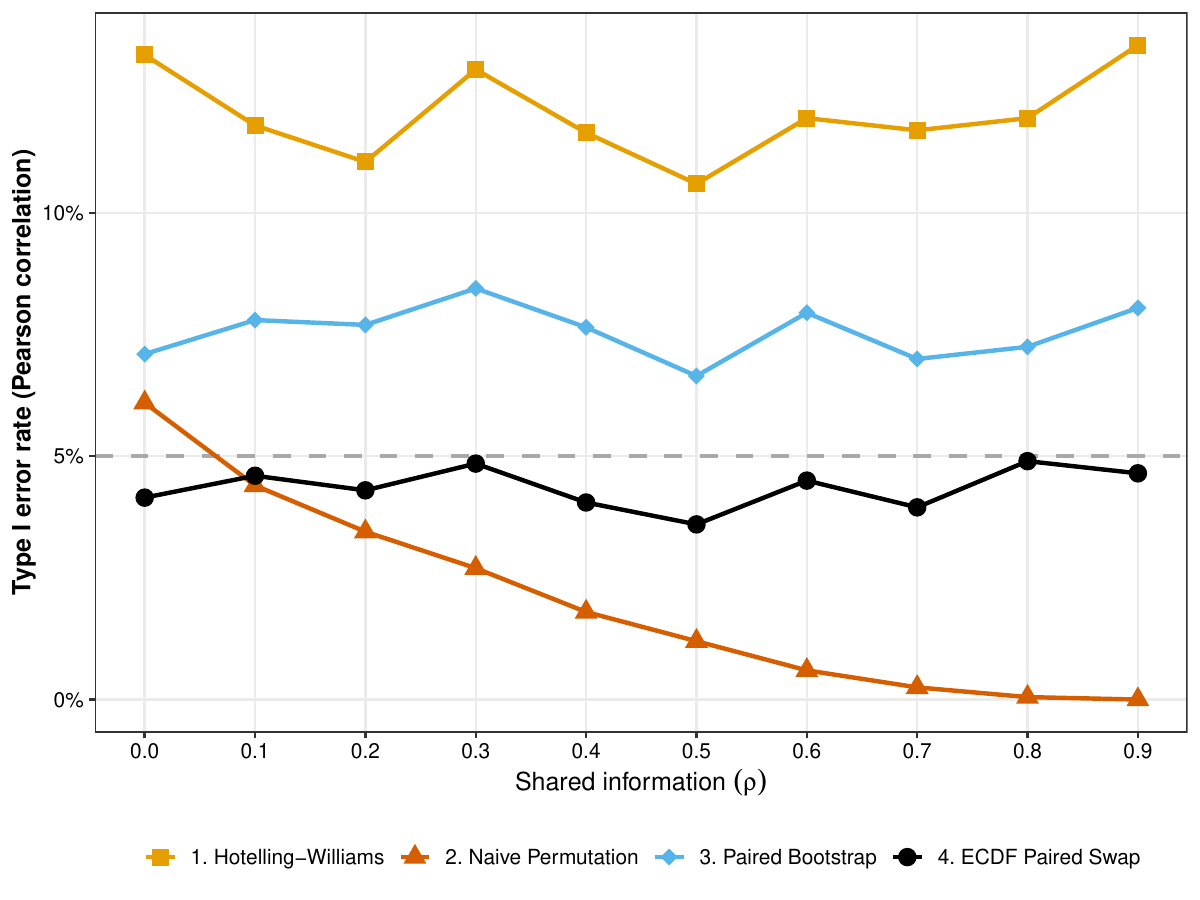} \includegraphics[width=7.5cm]{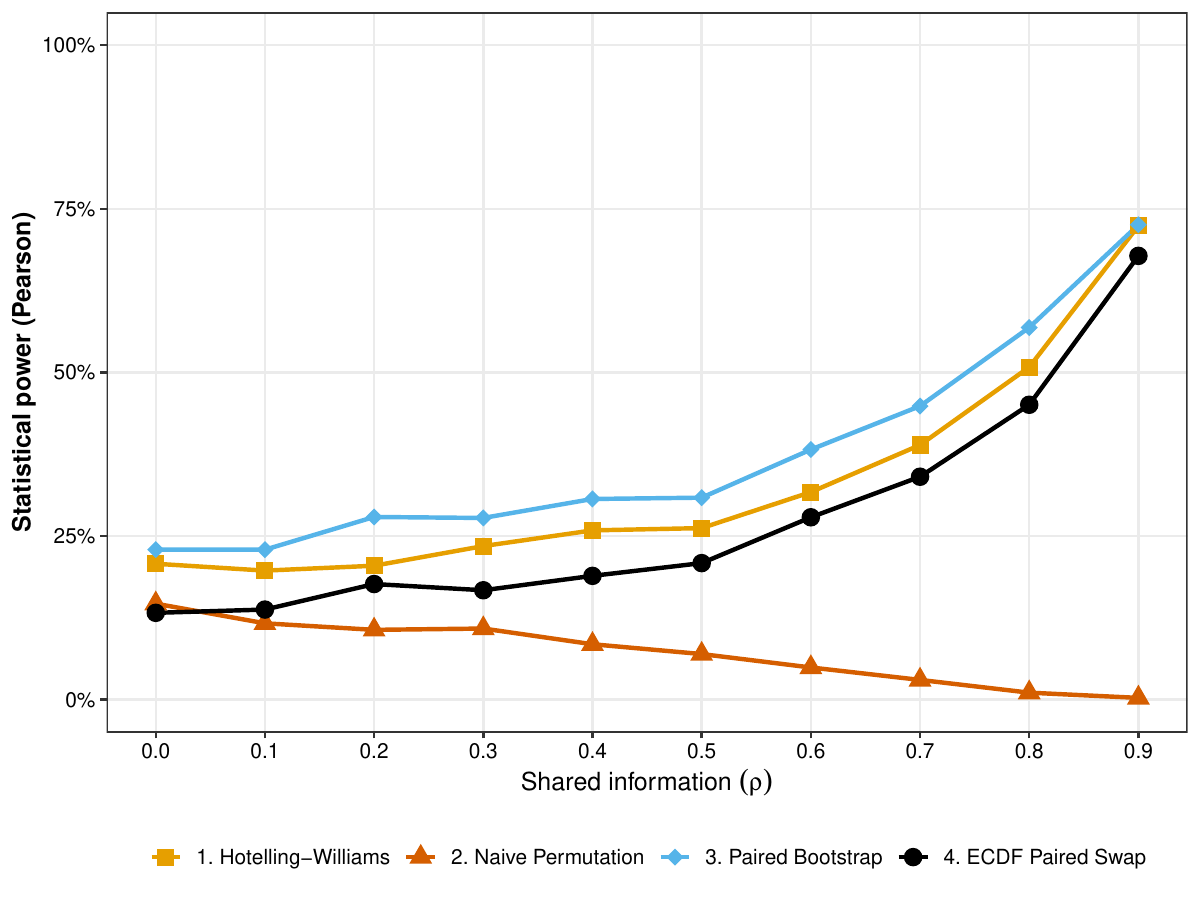}\\
    \includegraphics[width=7.5cm]{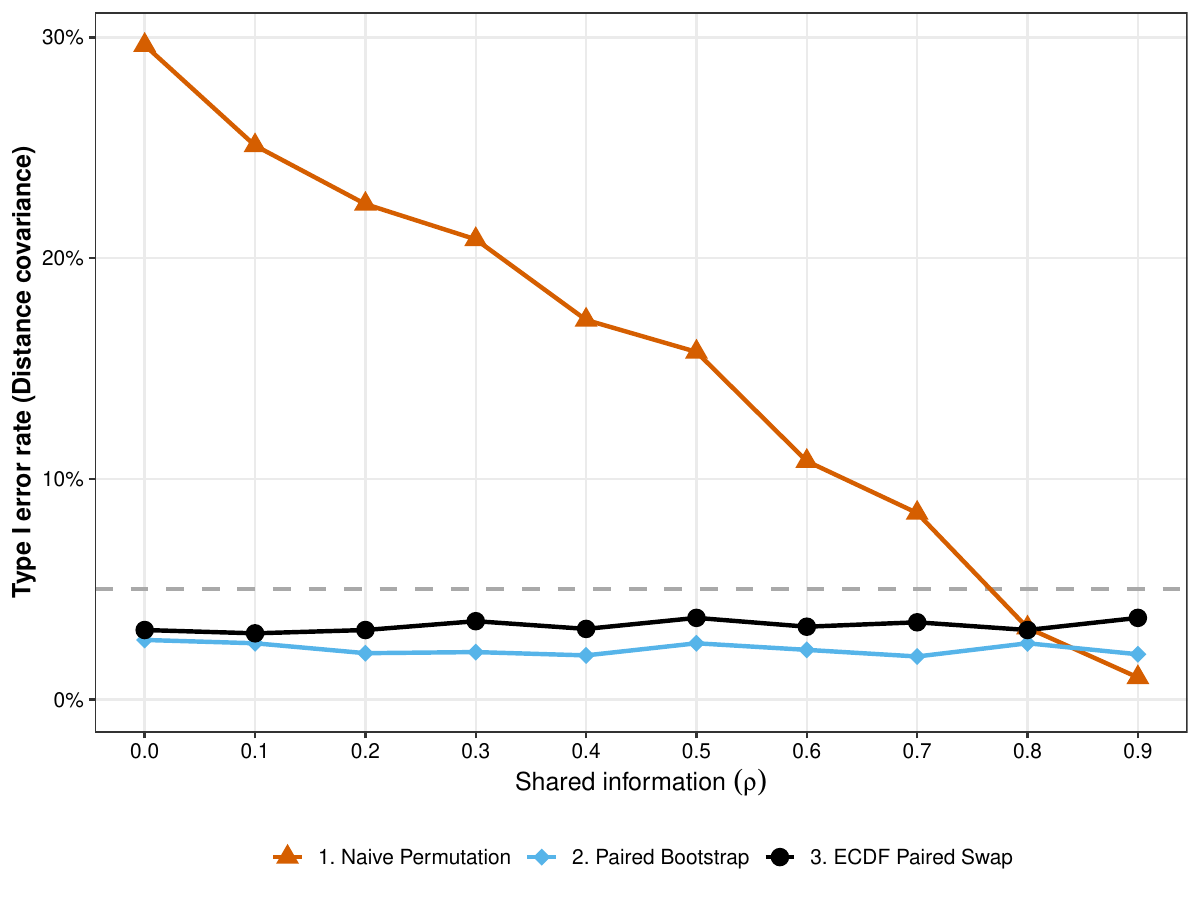} \includegraphics[width=7.5cm]{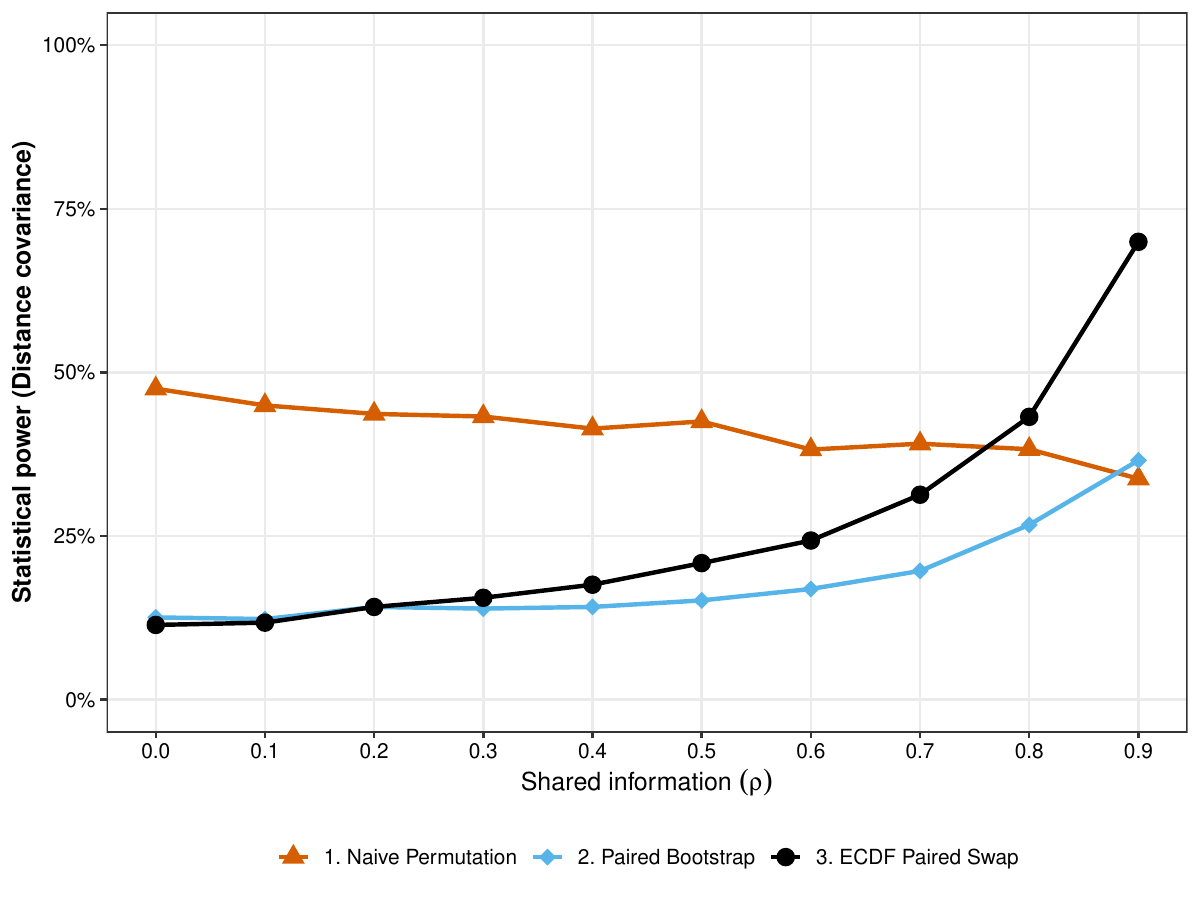}\\
    \caption{Continuous domain performance for Pearson's $r$ and Distance Covariance (dcov) across varying degrees of shared information $\rho$ at $N=30$. The top panels display the Type I error and Power for Pearson's $r$, while the bottom panels display the results for dcov. The dashed line indicates the nominal $\alpha = 0.05$ level.}
    \label{fig:continuous_linear_dcov}
\end{figure}

Figure \ref{fig:continuous_linear_dcov} illustrates the critical failures of classical methods in the presence of heavy tails. For strictly linear evaluations (Pearson), the Hotelling-Williams asymptotic test is catastrophically liberal, rejecting $H_0$ in $11-13\%$ of cases due to the severe variance inflation caused by the log-normal target $Y$. The naive permutation test exhibits a classic dependence breakdown: as the shared information $\rho$ increases, it becomes pathologically conservative, reaching a $0\%$ rejection rate and entirely destroying statistical power. While the paired bootstrap preserves the dependence structure, its evaluation of variance around $H_1$ under heavy tails makes it systematically liberal ($\sim 8\%$ Type I error). Our ECDF Paired Swap tightly holds the nominal $5\%$ level across the entire $\rho$ spectrum while delivering valid statistical power.

For non-linear detection using Distance Covariance, the paired bootstrap completely collapses. Because resampling with replacement generates localized ties, it artificially compresses the Euclidean metric space, resulting in severe conservativeness ($\sim 2\%$ Type I error) and paralyzing statistical power ($\sim 12-36\%$). In contrast, the ECDF Paired Swap guarantees the retention of all unique observations. By strictly mapping ranks through the empirical copula without compressing the topological space, our exact test successfully recovers the power profile of distance covariance, significantly outperforming the bootstrap.

\subsubsection{Continuous Evaluation: Kendall's Rank Correlation}
\begin{figure}[htpb]
    \centering
    \includegraphics[width=7.5cm]{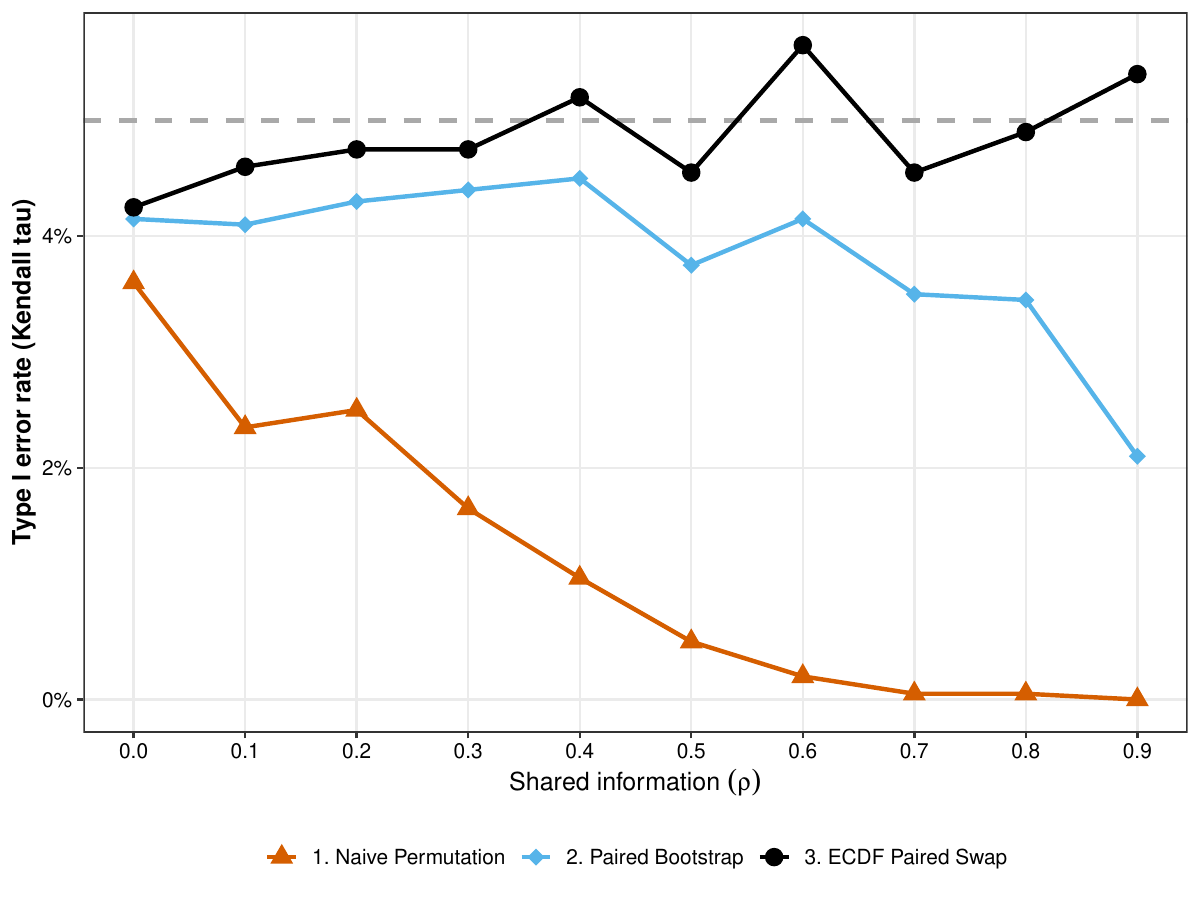} \includegraphics[width=7.5cm]{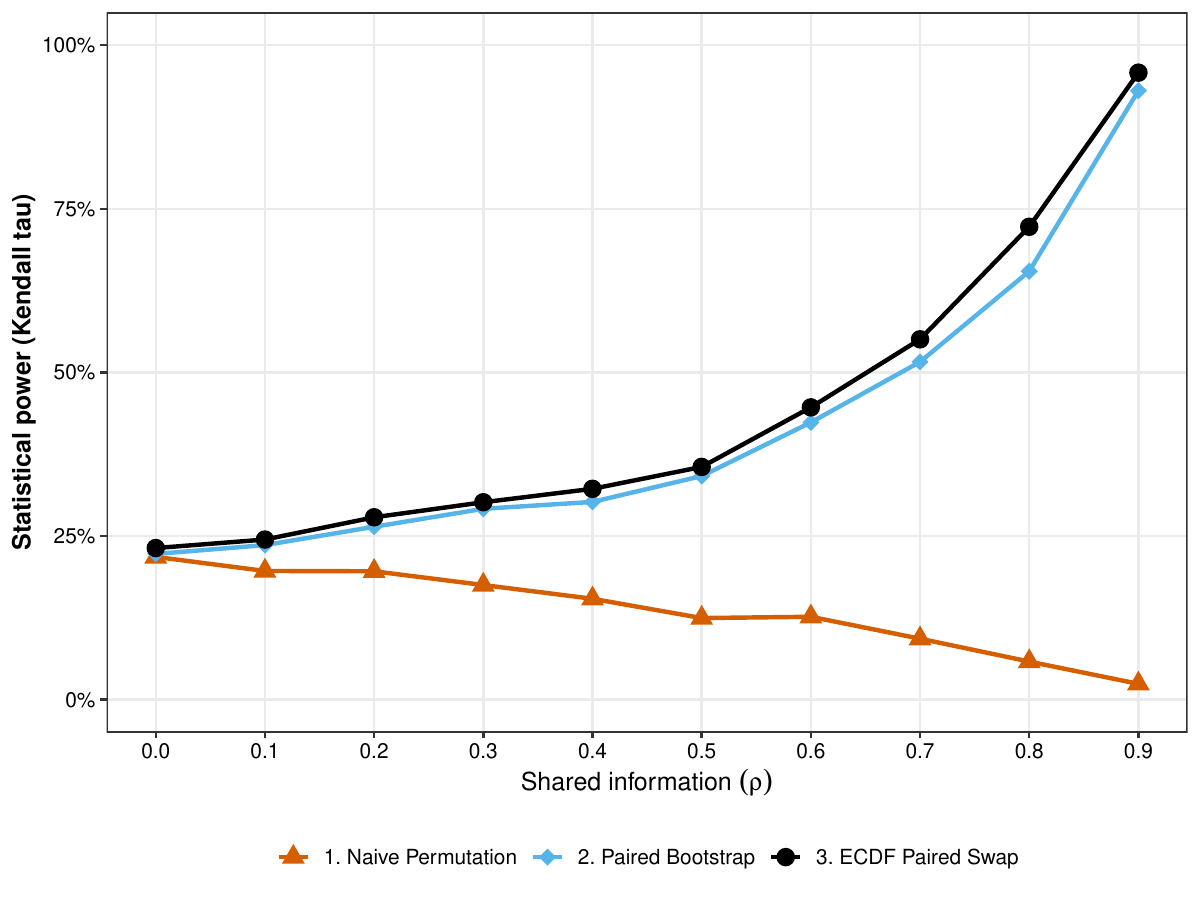}\\
    \includegraphics[width=7.5cm]{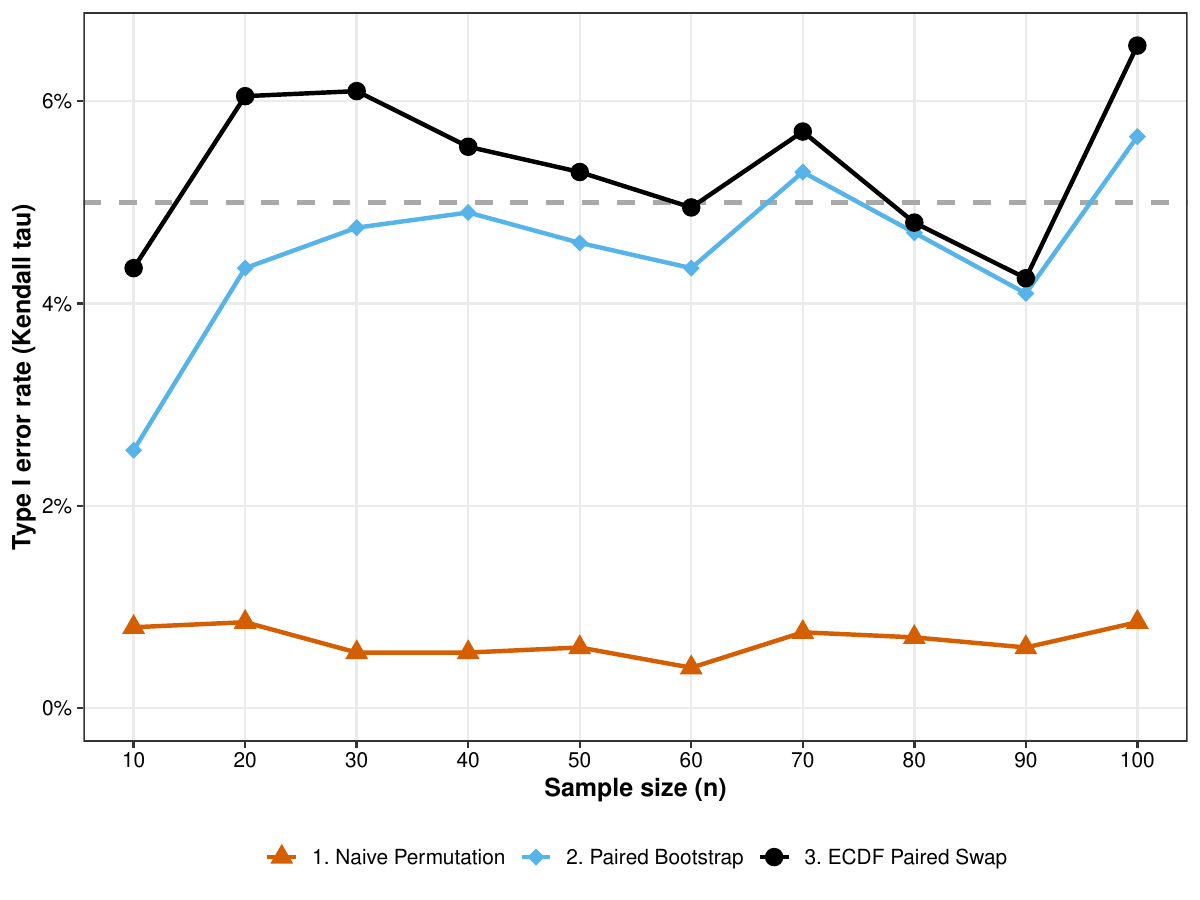} \includegraphics[width=7.5cm]{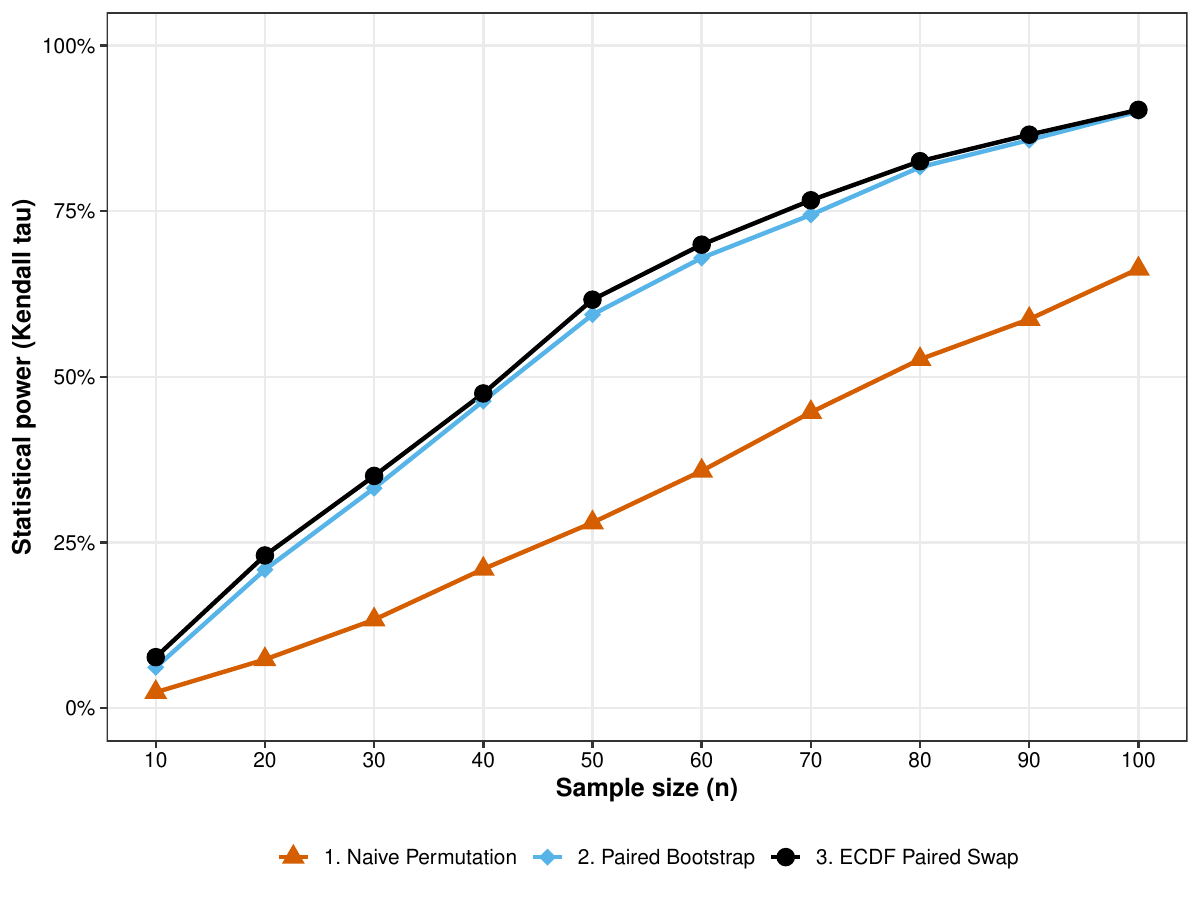}\\
    \caption{Continuous domain performance for Kendall's $\tau$. The top panels display Type I error and Power as a function of shared information $\rho$ ($N=30$). The bottom panels illustrate performance as a function of sample size $N$ ($\rho=0.5$).}
    \label{fig:continuous_kendall}
\end{figure}

Figure \ref{fig:continuous_kendall} highlights the behavior of the rank-based estimator Kendall's $\tau$. Unlike Pearson's $r$, rank transformations naturally neutralize the heavy tails of $Y$. At first glance, the paired bootstrap appears to perform reasonably well; however, this empirical stability is deceptive, likely resulting from competing distortions balancing each other out. Because the bootstrap resamples from the empirical distribution, it evaluates variance around the alternative hypothesis ($H_1$) rather than the true null ($H_0$). Depending on the exact data pathology, this $H_1$-centric variance estimation can induce unpredictable shifts toward either liberality or conservativeness. As demonstrated in the panels, for Kendall's $\tau$, this mechanism systematically drives the bootstrap to become overly conservative at small sample sizes ($N$) and under high shared information ($\rho$), resulting in a noticeable loss of statistical power. In contrast, the ECDF Paired Swap evaluates the exact null distribution, consistently holding the nominal significance level and dominating the bootstrap in power across all sample sizes and correlation states. The naive permutation, as mathematically expected, systematically collapses in both Type I validity and power as the dependence $\rho$ increases.

\subsubsection{Discrete Evaluation: Mutual Information}
\begin{figure}[htpb]
    \centering
    \includegraphics[width=7.5cm]{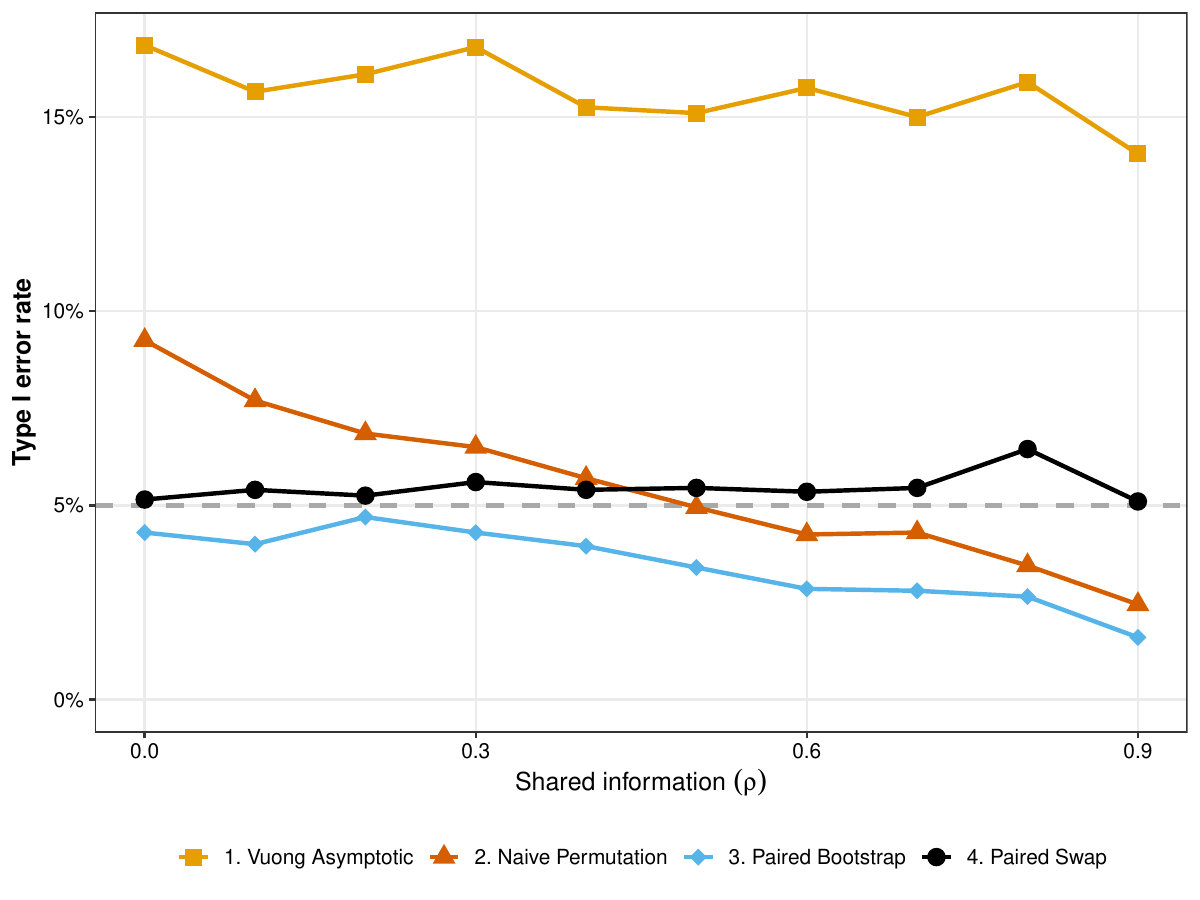} \includegraphics[width=7.5cm]{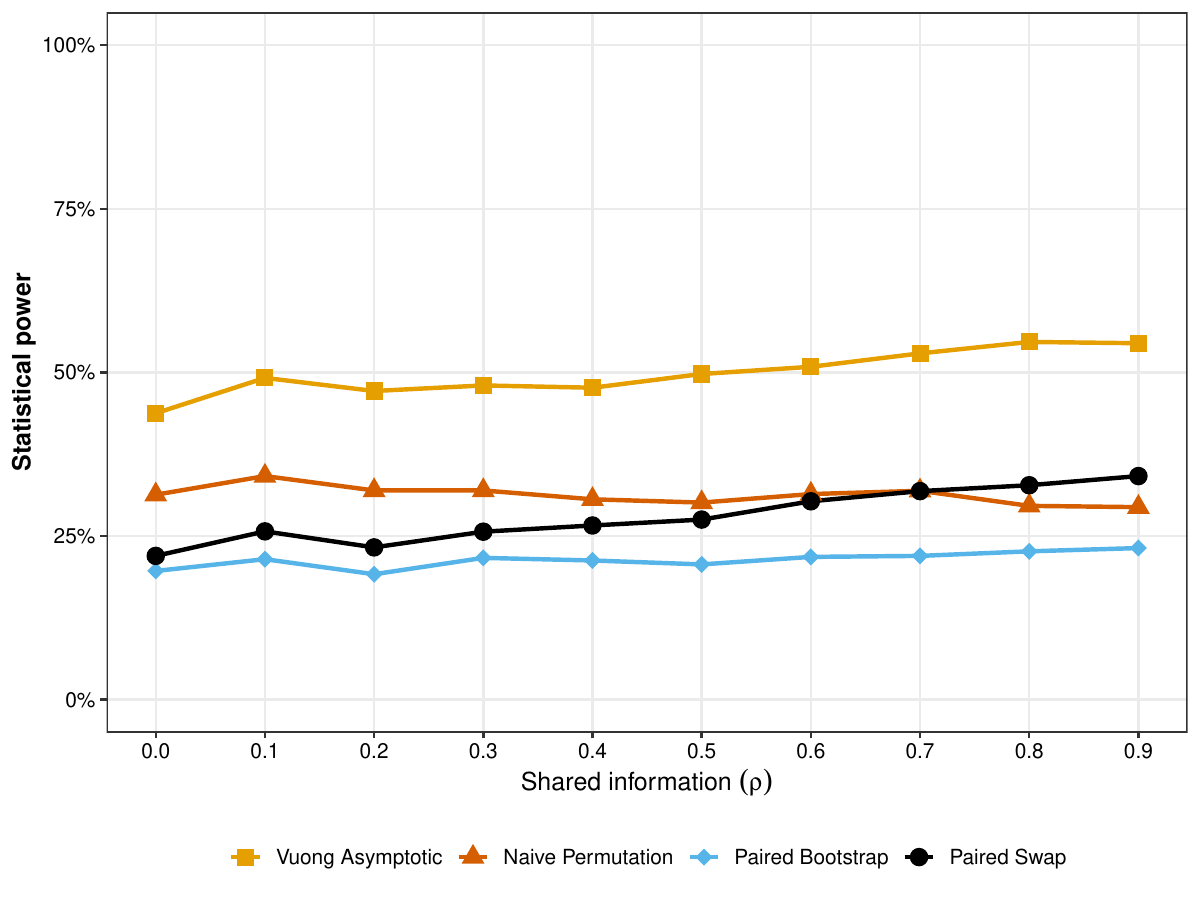}\\
    \includegraphics[width=7.5cm]{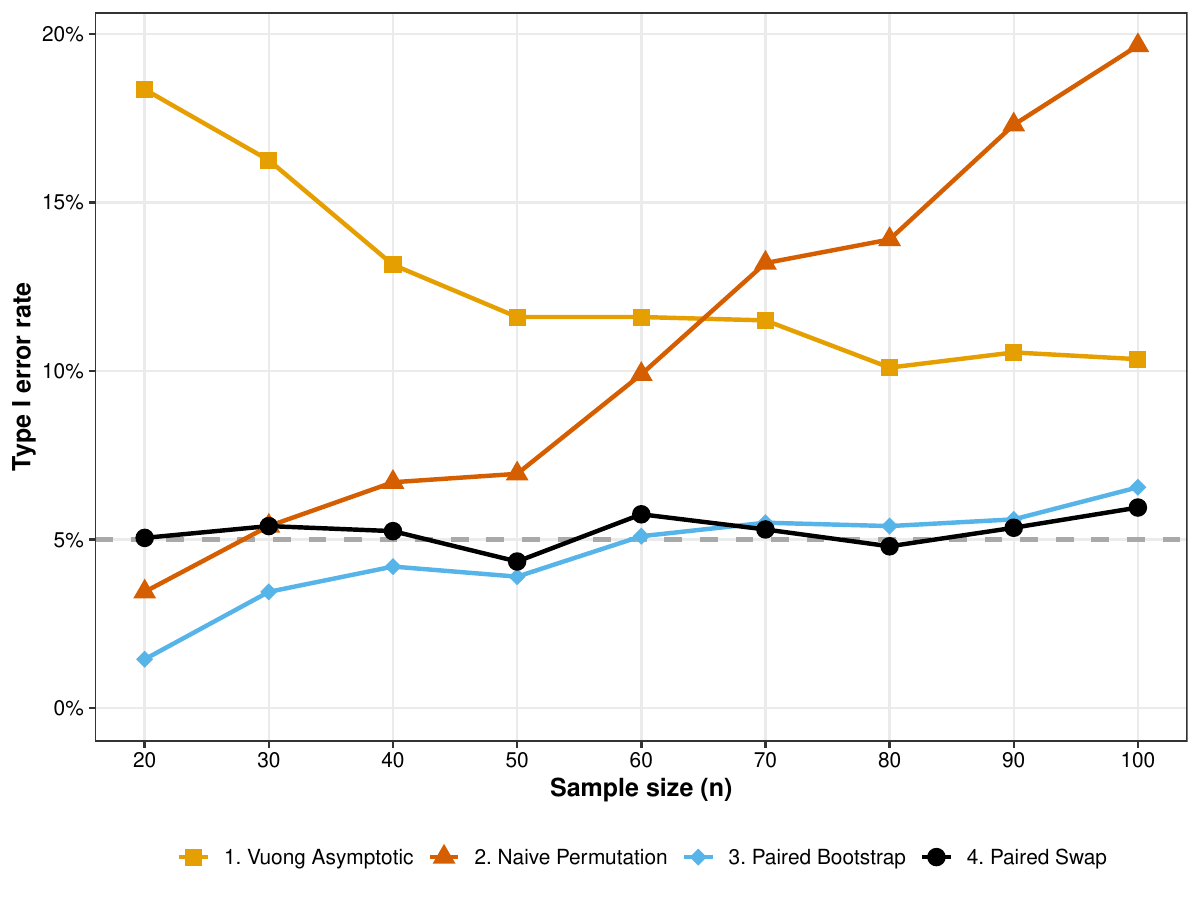} \includegraphics[width=7.5cm]{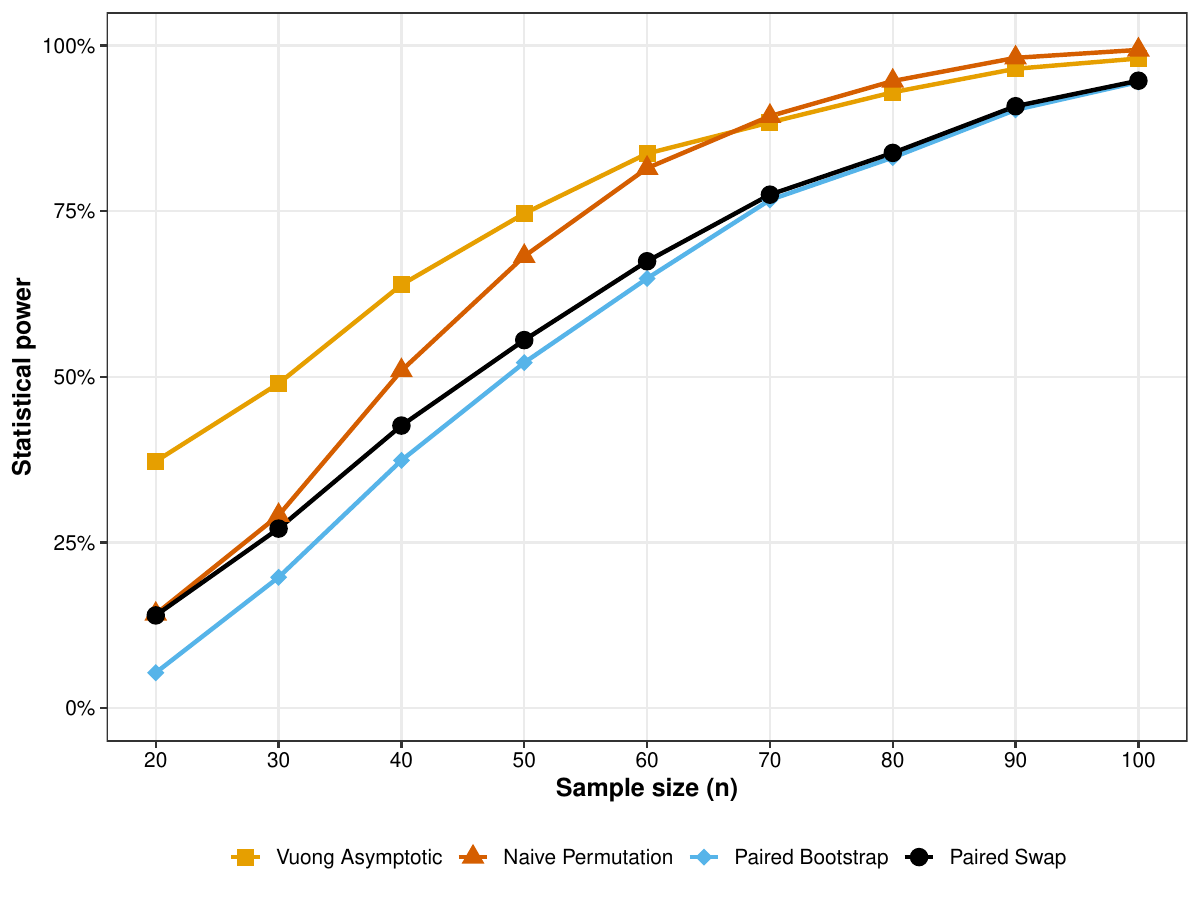}\\
    \caption{Categorical domain performance using Mutual Information. The top panels show Type I error and Power against varying shared information $\rho$ ($N=30$). The bottom panels display the dependency on sample size $N$ ($\rho=0.5$).}
    \label{fig:discrete}
\end{figure}

The discrete simulation results (Figure \ref{fig:discrete}) empirically confirm the breakdown of asymptotic and bootstrap methods under data sparsity. Vuong's closeness test is severely compromised; due to the presence of empty cells in the finite contingency tables ($N \le 100$), the likelihood ratios diverge, resulting in a massively inflated Type I error rate ($10-18\%$). Consequently, its apparent high power is an invalid artifact of this inflation.

The paired bootstrap, while conceptually sound, suffers from sparsity-induced entropy compression. By omitting rare categories during resampling, it structurally deflates the Mutual Information space, yielding a pathologically conservative test (dropping as low as $1.4\%$ Type I error at $N=20$). This structural bias translates directly to a severe suppression of statistical power.

Our proposed Paired Swap Permutation test demonstrates absolute robustness in this categorical environment. Because the within-subject symmetric swap mechanically preserves both the joint occurrence frequencies and the full vocabulary of categories, it is the only method that flawlessly maintains the $5\%$ significance level across all sample sizes and noise structures. Furthermore, it achieves the highest valid statistical power among all non-parametric alternatives, definitively proving its superiority for categorical model selection.

\section{Empirical Application: Explanatory Power in Linguistics}
\label{sec:case_study}

To illustrate the practical implications and the strict necessity of our methodology, we apply the \textit{Paired Swap Permutation Test} to a high-dimensional problem in morphology and corpus linguistics: predicting the semantic classification of Italian left-headed N+N (noun-noun) compounds.

\subsection{Data and Problem Complexity}
In this application, the target variable $Y$ represents the semantic classification (type) of the compound. The competing predictors are the constituent lemmas: $X_1$ denotes the first noun (N1) and $X_2$ denotes the second noun (N2). The central linguistic question is whether the first or the second constituent is more likely to serve as the trigger of the compound interpretation, thereby providing stronger predictive power with respect to the compound's classification. 

Data collected for the purpose of this study represent 5,798 lemmatized Italian left-headed N+N compounds, filtered out from Google n-gram frequency lists (Version~3;\citealt{google_google_2020}). The dataset was hand-cleaned and checked by a native speaker. Subsequently, each compound was assigned to one of three types—VNX (verbal-nexus), ATAP (attributive-appositive), and GRD (grounding)—which represent three fundamental macro-types of asymmetric compounds as defined in the influential Scalise-Bisetto classification \citep{scalise_classification_2009}. The choice of these categories is far from arbitrary, as it implies different initial expectations with respect to the potential predictors $X_1$ and $X_2$. Indeed, three main expectations may be inferred from the literature. First, $X_2$ is commonly assumed to be a good predictor of the ATAP relationship \citep{baroni_sulla_2009}, although one study \citep{radimsky_expanding_2025} suggests that in some cases $X_1$ may be equally informative. Second, $X_1$ should be a good predictor of the VNX relationship because VNX compounds require event nouns in the N1 position. Third, we have relatively low expectations regarding the predictive power of either $X_1$ or $X_2$ in GRD compounds: previous research suggests a potentially higher predictive power of $X_1$ \citep{baroni_sulla_2009}, whereas similar studies on French data have shown a rather balanced family-size effect of the two constituents in GRD N+N compounds \citep{radimsky_paradigmatic_2020}. From the quantitative viewpoint, ATAP and GRD compounds have a balanced representation in our sample (2,236 and 2,214 items, respectively), while the number of VNX compounds is slightly lower (1,348 items). Overall, these compounds contain 1,354 and 1,694 unique lemmas in the $X_1$ or $X_2$ position, respectively.

This dataset epitomizes the pathological conditions under which classical statistical inference breaks down. First, the predictors exhibit extreme cardinality and data sparsity; the vocabularies of $X_1$ and $X_2$ contain thousands of unique lemmas, the vast majority of which follow a Zipfian distribution with extremely low type frequencies in compounds. Second, both constituents are highly dependent, strictly constrained by grammatical and semantic co-occurrence rules. Finally, the collective explanatory power of the predictors is exceptionally high. Out of the total entropy of the compound type ($H(Y) = 1.550$ bits), the joint mutual information of both constituents explains nearly everything ($I(X_1, X_2; Y) = 1.549$ bits), leaving a negligible residual entropy of just $H(Y|X_1, X_2) = 0.00069$ bits. 

Although knowing both constituents almost deterministically yields the compound classification, there is still a non-negligible amount of independent information conveyed by each constituent. The individual conditional entropies of N1 and N2 are $H(Y|X_1) = 0.1851$ bits and $H(Y|X_2) = 0.3026$ bits, respectively. Equivalently, their individual mutual informations are $I(X_1; Y) = 1.365$ and $I(X_2; Y) = 1.248$ bits. Furthermore, the conditional mutual informations, $I(X_1; Y | X_2) = 0.302$ and $I(X_2; Y | X_1) = 0.184$, emphasize that while the first constituent provides slightly more unique information, the shared informational overlap (1.064 bits) is substantial, representing 68.6\% of the total entropy $H(y)$.

If one attempts to analyze the predictive difference between $X_1$ and $X_2$ using Vuong's closeness test via multinomial logistic regression, the estimation inevitably collapses. The sheer number of parameters required to fit the unique lemmas  combined with zero-frequency contingency cells leads to perfect separation and massive overfitting. This renders the asymptotic log-likelihood ratios mathematically unstable and the resulting $p$-values completely unreliable.

\subsection{Application of the Paired Swap Test}
To rigorously evaluate the relative predictive strength without relying on unstable likelihood asymptotics or biased bootstrap resampling, we applied the categorical variant of the \textit{Paired Swap Permutation Test}. We utilized Mutual Information as our test metric, evaluating the null hypothesis $H_0: I(X_1; Y) = I(X_2; Y)$. Because $X_1$ and $X_2$ are drawn from the identical lexical state space (Italian nouns), they natively satisfy the structural requirements for the symmetric within-subject swap.

We executed the exact test using $B = 10,000$ permutations. The observed difference in Mutual Information was precisely evaluated as $\Delta I_{obs} = I(X_1; Y) - I(X_2; Y) = 0.1176$ bits. A higher Mutual Information directly corresponds to a greater explanatory power; thus, the positive difference definitively demonstrates that $X_1$ resolves more uncertainty regarding the semantic type $Y$ than $X_2$.

When plotted against the exact null distribution generated strictly under $H_0$, the observed test statistic fell extraordinarily far into the right tail, completely outside the bounds of the simulated permuted distribution space. Consequently, the test yielded an exact empirical $p$-value of effectively zero ($p < 0.0001$). 

This computational result provides definitive, non-parametric evidence that the first constituent lemma ($X_1$) possesses a statistically significantly greater explanatory power than the second constituent ($X_2$). By deploying the symmetric swap mechanism, we identified $X_1$ as the dominant trigger of the interpretation of Italian N+N compounds, achieving robust inference that is immune to dimensionality and sparsity artifacts that undermine standard analytical methods.

\section{Discussion and Conclusion}
\label{sec:conclusion}

In this paper, we introduced the \textit{Paired Swap Permutation Test}, an exact non-parametric methodology designed to evaluate the relative explanatory power of two dependent predictors regarding a target variable. We established a unified testing framework applicable to both categorical domains (via symmetric swapping for predictors sharing a state space) and continuous domains (via ECDF-mapped rank swapping). 

Our extensive simulation study highlighted the severe vulnerabilities of classical analytical approaches in applied settings. Under pathological data conditions—such as heavy-tailed distributions in continuous spaces or extreme data sparsity in discrete domains—asymptotic models like the Hotelling-Williams test and Vuong's closeness test systematically break down, yielding catastrophically liberal inference. Furthermore, we mathematically and empirically confirmed that naive permutation tests, which shuffle predictors independently, violate the exchangeability of dependent variables. 

Crucially, we exposed the theoretical and empirical limitations of the paired bootstrap. While often considered the non-parametric gold standard, the bootstrap fundamentally resamples from the empirical distribution, thereby evaluating the variance of the test statistic around the local alternative hypothesis ($H_1$) rather than the strict null ($H_0$). We demonstrated that this $H_1$-centric variance estimation renders the bootstrap a purely approximate method. Depending on the underlying data geometry, the chosen estimator, and the presence of resampling ties (which cause metric space compression and categorical omission), the bootstrap can become highly unstable. As observed in our simulations, it exhibits unpredictable shifts toward either severe conservativeness or artificial liberality, particularly under small sample sizes ($N$) and high shared information ($\rho$). While the bootstrap is asymptotically valid for large, well-behaved datasets without sparsity, its finite-sample behavior under high correlation renders it unreliable.

Consequently, the \textit{Paired Swap Permutation Test} emerges as the optimal inferential tool for comparing dependent predictors under complex, finite-sample conditions. By evaluating the test statistic strictly under the null hypothesis without compressing the empirical support, our method mathematically guarantees an exact test that flawlessly maintains the nominal significance level while maximizing valid statistical power. As demonstrated in our linguistic case study on Italian noun-noun compounds, the exact paired swap provides robust, reliable inference even in domains plagued by massive dimensionality, perfect separation, and near-zero residual entropy—pathological environments where all conventional asymptotic and resampling methods inherently fail.

\vspace{1em}
\noindent \textbf{Acknowledgements:}
This work was supported by the Czech Science Foundation (GAČR) under project no.~25-15350S, \textit{How do word-formation patterns emerge? An empirical diachronic analysis of Italian N+N compounds}, conducted at the Faculty of Arts, University of South Bohemia in České Budějovice.

\vspace{1em}
\noindent \textbf{Disclosure regarding AI use:} The authors utilized a generative AI assistant solely for language refinement and copy-editing to ensure standard academic English. All methodological innovations, statistical implementations, and data interpretations remain the independent work of the authors. The authors have reviewed and assume full responsibility for the content of the final manuscript.

\bibliographystyle{apalike}
\bibliography{Tomas_bibfile, Honza_bibfile}

@incollection{scalise_classification_2009,
	title = {The {Classification} of {Compounds}},
	isbn = {978-0-19-969572-0},
	url = {https://doi.org/10.1093/oxfordhb/9780199695720.013.0003},
	doi = {10.1093/oxfordhb/9780199695720.013.0003},
	booktitle = {The {Oxford} {Handbook} of {Compounding}},
	publisher = {Oxford University Press},
	author = {Scalise, Sergio and Bisetto, Antonietta},
	editor = {Lieber, Rochelle and Štekauer, Pavol},
	year = {2009},
	pages = {34--53},
}

@incollection{baroni_sulla_2009,
	title = {Sulla tipologia dei composti {N}+{N} in italiano : principi categoriali ed evidenza distribuzionale a confronto},
	shorttitle = {Sulla tipologia dei composti {N}+{N} in italiano},
	url = {https://doi.org/10.1400/137523},
	language = {ita},
	urldate = {2022-02-16},
	booktitle = {Linguistica e modelli tecnologici di ricerca : atti del {XL} {Congresso} internazionale di studi della {Società} di linguistica italiana ({SLI}) : {Vercelli}, 21-23 settembre 2006},
	publisher = {Bulzoni Editore},
	author = {Baroni, Marco and Guevara, Emiliano and Pirrelli, Vito},
	editor = {Ferrari, Giacomo and Benatti, Ruben and Mosca, Monica},
	year = {2009},
	pages = {73--96},
}

@article{radimsky_paradigmatic_2020,
	title = {A paradigmatic account of lexical innovation: the role of repeated components in {French} {N}+{N} compounds},
	volume = {12},
	copyright = {All rights reserved},
	url = {https://pasithee.library.upatras.gr/mmm/article/view/3251/3511},
	doi = {10.26220/mmm.3251},
	journal = {Mediterranean Morphology Meetings; Vol 12 (2019): Rules, patterns, schemas and analogy},
	author = {Radimský, Jan},
	year = {2020},
	pages = {77--91},
}

@article{radimsky_expanding_2025,
	title = {Expanding {Horizons}: {A} comprehensive insight into the evolution of {Italian} {Attributive}-{Appositive} {Noun}+{Noun} {Compounds} using {Google} {Books} {Data}},
	copyright = {Creative Commons Attribution-NonCommercial-NoDerivatives 4.0 International License (CC-BY-NC-ND)},
	issn = {2804-7397},
	shorttitle = {Expanding {Horizons}},
	url = {http://www.peren-revues.fr/lexique/index.php?id=1948},
	doi = {10.54563/lexique.1948},
	language = {fr},
	number = {Numéro spécial},
	urldate = {2025-05-02},
	journal = {Lexique},
	publisher = {Université de Lille},
	author = {Radimský, Jan and Micheli, M. Silvia},
	year = {2025},
	pages = {53--74},
}

@article{gagne_constituent_2009,
	title = {Constituent integration during the processing of compound words: {Does} it involve the use of relational structures?},
	volume = {60},
	copyright = {https://www.elsevier.com/tdm/userlicense/1.0/},
	issn = {0749596X},
	shorttitle = {Constituent integration during the processing of compound words},
	url = {https://linkinghub.elsevier.com/retrieve/pii/S0749596X08000673},
	doi = {10.1016/j.jml.2008.07.003},
	language = {en},
	number = {1},
	urldate = {2026-06-23},
	journal = {Journal of Memory and Language},
	author = {Gagné, Christina L. and Spalding, Thomas L.},
	month = jan,
	year = {2009},
	pages = {20--35},
}

@article{gagne_influence_1997,
	title = {Influence of thematic relations on the comprehension of modifier–noun combinations.},
	volume = {23},
	issn = {1939-1285, 0278-7393},
	url = {https://doi.apa.org/doi/10.1037/0278-7393.23.1.71},
	doi = {10.1037/0278-7393.23.1.71},
	language = {en},
	number = {1},
	urldate = {2026-06-23},
	journal = {Journal of Experimental Psychology: Learning, Memory, and Cognition},
	author = {Gagné, Christina L. and Shoben, Edward J.},
	month = jan,
	year = {1997},
	pages = {71--87},
}

@article{de_jong_morphological_2000,
	title = {The morphological family size effect and morphology},
	volume = {15},
	issn = {0169-0965, 1464-0732},
	url = {https://www.tandfonline.com/doi/full/10.1080/01690960050119625},
	doi = {10.1080/01690960050119625},
	language = {en},
	number = {4-5},
	urldate = {2026-06-23},
	journal = {Language and Cognitive Processes},
	author = {De Jong, Nivja H. and Schreuder, Robert and Harald Baayen, R.},
	month = aug,
	year = {2000},
	pages = {329--365},
}

@article{gunther_caoss_2021,
	title = {{CAOSS} and transcendence: {Modeling} role-dependent constituent meanings in compounds},
	volume = {33},
	issn = {1871-5656},
	url = {https://doi.org/10.1007/s11525-021-09386-6},
	doi = {10.1007/s11525-021-09386-6},
	number = {4},
	journal = {Morphology},
	author = {Günther, Fritz and Marelli, Marco},
	year = {2021},
	pages = {409--432},
}

@misc{google_google_2020,
	title = {Google {Books} {Ngram} {Corpus}, {Version} 3},
	url = {https://storage.googleapis.com/books/ngrams/books/datasetsv3.html},
	author = {{Google}},
	year = {2020},
}

@article{vuong1989likelihood,
  title={Likelihood ratio tests for model selection and non-nested hypotheses},
  author={Vuong, Quang H},
  journal={Econometrica: Journal of the Econometric Society},
  volume={57},
  number={2},
  pages={307--333},
  year={1989},
  publisher={JSTOR}
}

@article{szekely2007measuring,
  title={Measuring and testing dependence by correlation of distances},
  author={Sz{\'e}kely, G{\'a}bor J and Rizzo, Maria L and Bakirov, Nail K},
  journal={The Annals of Statistics},
  volume={35},
  number={6},
  pages={2769--2794},
  year={2007},
  publisher={Institute of Mathematical Statistics}
}

@article{ernst2004permutation,
  title={Permutation methods: a basis for exact inference},
  author={Ernst, Michael D},
  journal={Statistical Science},
  volume={19},
  number={4},
  pages={676--685},
  year={2004},
  publisher={Institute of Mathematical Statistics}
}

@book{pesarin2010permutation,
  title={Permutation tests for complex data: theory, applications and software},
  author={Pesarin, Fortunato and Salmaso, Luigi},
  year={2010},
  publisher={John Wiley \& Sons}
}

@book{efron1993introduction,
  title={An introduction to the bootstrap},
  author={Efron, Bradley and Tibshirani, Robert J},
  year={1993},
  publisher={CRC press}
}

@article{hotelling1940selection,
  title={The selection of variates for use in prediction with some comments on the general problem of nuisance parameters},
  author={Hotelling, Harold},
  journal={The Annals of Mathematical Statistics},
  volume={11},
  number={3},
  pages={271--283},
  year={1940},
  publisher={JSTOR}
}

@article{williams1959comparison,
  title={The comparison of regression variables},
  author={Williams, Evan J},
  journal={Journal of the Royal Statistical Society: Series B (Methodological)},
  volume={21},
  number={2},
  pages={396--399},
  year={1959},
  publisher={Wiley Online Library}
}

@article{steiger1980tests,
  title={Tests for comparing elements of a correlation matrix},
  author={Steiger, James H},
  journal={Psychological Bulletin},
  volume={87},
  number={2},
  pages={245--251},
  year={1980},
  publisher={American Psychological Association}
}

@article{strobl2008conditional,
  title={Conditional variable importance for random forests},
  author={Strobl, Carolin and Boulesteix, Anne-Laure and Kneib, Thomas and Augustin, Thomas and Zeileis, Achim},
  journal={BMC bioinformatics},
  volume={9},
  number={1},
  pages={307},
  year={2008},
  publisher={BioMed Central}
}

@article{abney2015permutation,
  title={Permutation testing in the presence of polygenic variation},
  author={Abney, Mark},
  journal={The American Journal of Human Genetics},
  volume={96},
  number={3},
  pages={412--428},
  year={2015},
  publisher={Elsevier}
}

\end{document}